\documentclass{vgtc}                          

\ifpdf                             
  \pdfoutput=1\relax               
  \usepackage{graphicx}               
  \DeclareGraphicsExtensions{.pdf,.png,.jpg,.jpeg} 
\else                             
  \usepackage{graphicx}              
  \DeclareGraphicsExtensions{.eps}    
\fi

\graphicspath{{figures/}{pictures/}{images/}{./}}

\usepackage{microtype}                 
\PassOptionsToPackage{warn}{textcomp}  
\usepackage{textcomp}                 
\usepackage{mathptmx}                  
\usepackage{times}                     
         
\usepackage{cite}                      
\usepackage{tabu}                      
\usepackage{booktabs}                  
\usepackage{xspace}
\usepackage{enumitem}

\usepackage{hyperref}

\onlineid{0}

\vgtccategory{Research}

\vgtcinsertpkg

\title{\tool{}: Interactive Visual Exploration of Literature in Browsers }

\author{
    Kevin Li$^\ast$, 
    Haoyang Yang$^\ast$,
    Anish Upadhayay$^\ast$, 
    Zhiyan Zhou$^\ast$,
    Jon Saad-Falcon$^\ast$,
    Duen Horng (Polo) Chau\thanks{
            Georgia Institute of Technology.\newline 
            \{kevin.li,alexanderyang,aupadhayay3,zzhou406,jonsaadfalcon,
            polo\}@gatech.edu
        }$\,\,\,$
}

\teaser{
  \centering
  \includegraphics[width=\linewidth]{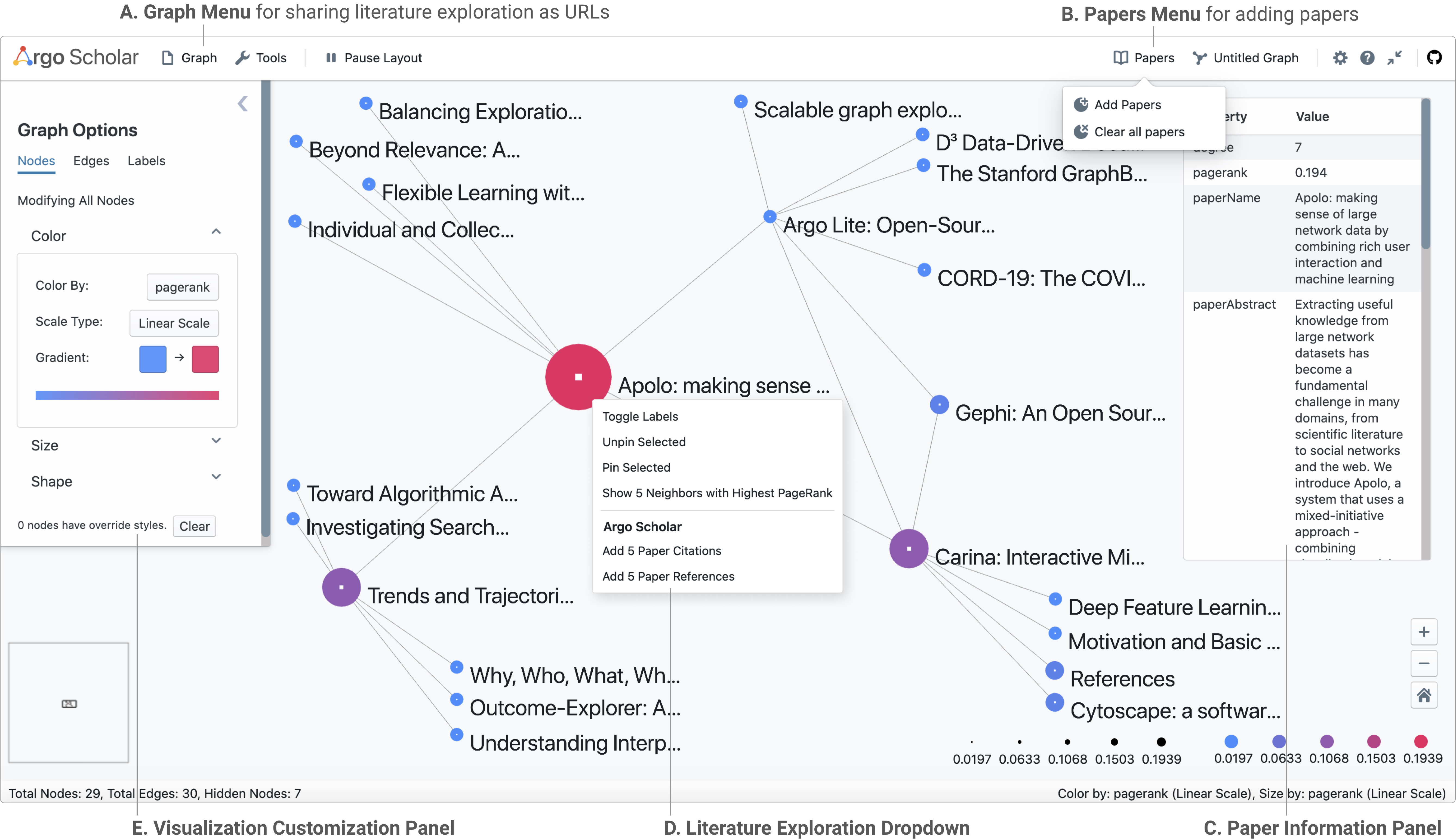}
  \caption{\tool{} visualizing a literature network of papers (nodes) and citations (edges) around the \textit{Apolo} article. 
  \textbf{A. Graph Menu} for saving and sharing user's literature exploration snapshots as URLs.
  \textbf{B. Papers Menu} for creating new exploration and adding papers via their Semantic Scholar CorpusIDs.
  \textbf{C. Paper Information Panel} shows key information (e.g., paper title, abstract, citation count). 
  \textbf{D. Literature Exploration Dropdown} for incrementally adding more connected papers that are citing or being cited. 
  \textbf{E. Visualization Customization Panel} for modifying visual properties of the graph of papers (e.g., color, size, labeling).
  }
  \label{fig:teaser}
  \bigskip
}

\abstract{
Discovering and making sense of relevant research literature is fundamental to becoming knowledgeable in any scientific discipline.
Visualization can aid this process; 
however, existing tools' adoption and impact have often been constrained, such as by their reliance on small curated paper datasets that quickly become outdated or a lack of support for personalized exploration.
We introduce \tool{}, an open-source, web-based visualization tool for interactive exploration of literature and easy sharing of exploration results.
\tool{} queries and visualizes Semantic Scholar's live data of almost 200 million papers, enabling users to generate personalized literature exploration results in real-time
through flexible, incremental exploration, a common and effective method for researchers to discover relevant work.
Our tool allows users to easily share their literature exploration results as a URL or web-embedded IFrame application.
\tool{} is open-sourced and available at  \textcolor{linkColor}{\url{https://poloclub.github.io/argo-scholar/}}. 
} 

\CCScatlist{
  \CCScatTwelve{Human-centered computing}{Visu\-al\-iza\-tion}{Visu\-al\-iza\-tion systems and tools}{};
}

\newcommand{\tool}[0]{\textsc{Argo Scholar}\xspace{}}

\definecolor{linkColor}{RGB}{6,125,233}

\begin{document}

\firstsection{Introduction}

\maketitle

Exploring and making sense of relevant research literature is fundamental to any scientific discipline. 
While visualization tools have been proposed to aid this critical task, their adoption and impact have often been constrained by their reliance on manually-curated paper datasets that were often small and outdated subsets of the available literature \cite{Kairam2015RefineryVE,3Chau2011ApoloMS}.
A recent visual tool called Connected Papers aims to automate the discovery of relevant work for users, but it shows all users the same visualization for a given paper and does not support incremental exploration or associative browsing, which are common effective strategies for discovering relevant work \cite{1PirolliTheSP,Kairam2015RefineryVE,2Pienta2015ScalableGE} (e.g., starting with a handful familiar or influential papers and branching out over time).
Addressing the common literature exploration needs shared by researchers, our ongoing work makes the following contributions:

\begin{enumerate}[itemsep=2mm, topsep=2mm, parsep=1mm, leftmargin=5mm]     
    \item \textbf{\tool{}, an open-source, web-based visualization tool for interactive exploration of literature and easy sharing of exploration results} (\autoref{fig:teaser}).
    Anyone can access \tool{} using their web browser on desktops and mobile devices without the need for any software installation. \tool{} is open-sourced under the permissive MIT license\footnote{\url{https://github.com/poloclub/argo-scholar}} and is available at the public link: \textcolor{linkColor}{\url{https://poloclub.github.io/argo-scholar/}}.

    \item \textbf{Incremental, flexible literature exploration with up-to-date data}. 
    \tool{} connects to the live data of Semantic Scholar \cite{5Lo2020S2ORCTS} via its API, gaining access to its 194 million indexed papers (as of June 2021).
    \tool{}'s ability to query and visualize up-to-date data in real-time overcomes major limitations faced by prior tools proposed for literature sensemaking that relied on manually-curated paper datasets that were often small and outdated subsets of available literature \cite{3Chau2011ApoloMS,Kairam2015RefineryVE}.
    Users can add papers indexed by Semantic Scholar by entering the papers' unique CorpusIDs (\autoref{fig:teaser}B). 
    And for each paper, 
    users can incrementally explore the connected papers that cite it or are cited,
    generating personalized literature networks for each user (\autoref{fig:teaser}D), unlike existing tools such as Connected Papers\footnote{\url{https://www.connectedpapers.com}} that show all users the same visualization for a given paper. 

    \item \textbf{Easy sharing of literature exploration results}. \tool{} allows users to share their literature networks and visualization as a URL or IFrame web-embedded application (\autoref{fig:teaser}A). 
    By accessing a shared link, 
    researchers build upon existing exploration progress to discover more relevant papers.
\end{enumerate}

\section{\tool{}'s System Design}
Developed using modern web technologies,
\tool{}'s interface provides users with new functionalities beneficial for literature sensemaking (e.g., incremental exploration, live data connection to Semantic Scholar) 
while inheriting the core graph visualization capabilities of Argo Lite \cite{4Li2020ArgoLO} (e.g., React for Blueprint for UI, MobX state management, Three.js WebGL graph rendering).

\vspace{\medskipamount}
\noindent
\textbf{Displaying Paper Relations.} 
\tool{}'s main view displays a citation network as a graph of papers (nodes) and citations (edges). 
Users can add any paper indexed by Semantic Scholar \cite{5Lo2020S2ORCTS} (\url{https://www.semanticscholar.org}) by entering the paper's unique CorpusID via the \textit{Papers Menu} (\autoref{fig:teaser}B).
From there, the user can branch out from currently visualized papers by exploring their citations and references or input new, potentially unrelated, papers.
Directed edges point from a citing paper to the cited; 
edge directions can be toggled.
A variety of visual properties of the nodes, edges, and node labels can be customized via the \textit{Customization Panel} (\autoref{fig:teaser}E).
For example, nodes can be colored and sized based on paper attributes such as citation count, with the ranges of both being user-adjustable. 

\vspace{\medskipamount}
\noindent

\textbf{Incremental Exploration.} 
Making sense of literature through incremental paper exploration or associative browsing is a common, effective method for researchers to discover relevant work \cite{Kairam2015RefineryVE,3Chau2011ApoloMS}.
Specifically, they may start with a handful of 
(familiar) papers, then gradually build their network of related papers through exploring the references and citations of 
their growing network. 
\tool{} supports this literature sensemaking strategy;
users can iteratively add 5 reference or citation papers for a paper\footnote{Semantic Scholar supports 100 requests per 5 minutes per IP address} via its \textit{Exploration Dropdown} (Fig~\ref{fig:teaser}D).
The paper is connected to the added papers by directed edges that encode the citation relationships previously described, and
the added papers are aligned vertically to improve label readability.  
Network data, such as the degree, pagerank, size, and color of existing nodes, are updated accordingly. 

\vspace{\medskipamount}
\noindent
\textbf{Learning More About Each Paper.} The \textit{Paper Information Panel} (Fig~\ref{fig:teaser}C) of \tool{} shows essential paper information for a highlighted or hovered-over node,
e.g., paper title, abstract, authors, citation count, publication venue, publication year, and the URL to its Semantic Scholar page (for paper PDF, figures, etc.). 

\vspace{\medskipamount}
\noindent
\textbf{Saving and Sharing Explorations Across Platforms.} \tool{} allows users to save and share their literature exploration as a \textit{snapshot} JSON file that stores the information of all explored papers, their connections, and visualization customizations via the \textit{Graph Menu} (Fig~\ref{fig:teaser}A).

Users have two options when saving and sharing their exploration snapshots: locally on their device or in the cloud. If users opt for the first option, they will download the JSON file and then upload it to \tool{} whenever they wish to resume their sensemaking process. Or, users can save their snapshots to \tool{}'s server for free, 
enabling them to share their literature network as a custom URL, bypassing the need to send JSON files. Saving the snapshot to \tool{}'s servers also grants the ability to embed the literature network as either an HTML IFrame or a Jupyter Notebook IFrame. \tool{} runs on all modern web browsers (e.g., Chrome, Safari, Firefox, Edge), broadening its access.

\section{Conclusion and Ongoing Work.} 
\tool{} is an 
open-source, web-based visualization tool for interactive exploration of literature and easy sharing of exploration results.
\tool{} queries and visualizes Semantic Scholar’s live data, enabling users to generate personalized literature exploration results in real-time through flexible, incremental exploration.
Anyone can access our tool using their web browser on desktops and mobile devices without the need for any software installation.

We plan to improve the user experience of \tool{} by 
enabling users to directly search and show paper results in \tool{}, bypassing the current need for 
locating the paper's CorpusID. This tighter connection with Semantic Scholar would allow us to enhance the usability and user-friendliness of our visualization tool. Additionally, we plan to improve the ordering of adding cited papers and referenced papers by implementing the ability to add them based on metrics such as citation count, recency, or relevance. These improvements would add more flexibility to the sensemaking process as users can further customize the way they explore literature.
We also plan to evaluate \tool{} through user studies with both beginning and seasoned researchers to evaluate \tool{}'s usability and its ability to help with literature sensemaking in the long term.
As \tool{} improves its functionalities, 
we look forward to more students, researchers, and practitioners adopting \tool{} to discover relevant research and make sense of the literature across disciplines.

\bibliographystyle{unsrt}

\bibliography{template}
\end{document}